\documentclass[10pt,conference]{IEEEtran}
	\IEEEoverridecommandlockouts
	\let\labelindent\relax
	\usepackage{algorithmic, algorithm2e, color}
	\usepackage{graphicx, subfigure, epstopdf}
	\usepackage{amsmath, amssymb, mathtools}
	\usepackage{graphicx, subfigure, epstopdf}
	\usepackage{exscale,relsize}
	\usepackage{amsfonts}
	\usepackage{mdwlist}

	\usepackage{array} 
	\usepackage{url}
	\usepackage{graphicx}
	\usepackage{subfigure}
	\usepackage{listings}
	\usepackage{dsfont}
	\usepackage{enumitem}
	\usepackage{paralist}

	\usepackage{times}

	\title{The Effect of Visual Noise on \\ The Completion of Security Critical Tasks * \thanks{*This work originally appeared in USEC 2015}}

	\author{Tyler Kaczmarek\\ UC Irvine \\ tkaczmar@uci.edu \and 
	Alfred Kobsa \\ UC Irvine \\ kobsa@uci.edu 
	\and  Robert Sy**\thanks{**this author's contributions were made while affiliated with UC Irvine} \\ DHS \\ robertsyproductions@yahoo.com 		\and 
	Gene Tsudik\\ UC Irvine \\ gene.tsudik@uci.edu} 

\begin{document}
	\IEEEoverridecommandlockouts
	\makeatletter\def\@IEEEpubidpullup{9\baselineskip}\makeatother
	\IEEEpubid{\parbox{\columnwidth}{ Permission to freely reproduce all or part
	    of this paper for noncommercial purposes is granted provided that
	    copies bear this notice and the full citation on the first
	    page. Reproduction for commercial purposes is strictly prohibited
	    without the prior written consent of the Internet Society, the
	    first-named author (for reproduction of an entire paper only), and
	    the author's employer if the paper was prepared within the scope
	    of employment.  \\ 
	    USEC '15, 8 February 2015, San Diego, CA, USA\\
	    Copyright 2015 Internet Society, ISBN 1-891562-40-1\\
	    http://dx.doi.org/10.14722/usec.2015.23014
	}
\hspace{\columnsep}\makebox[\columnwidth]{}}
	
	\maketitle

	\begin{abstract}
	User errors while performing security-critical tasks can lead to undesirable or even 
	disastrous consequences. One major factor influencing mistakes and failures is 
	complexity of such tasks, which has been studied extensively in prior research.
	Another important issue which hardly received any attention is the impact of both 
	accidental and intended distractions on users performing security-critical tasks. 
	In particular, it is unclear whether, and to what extent, unexpected sensory cues 
	(e.g., auditory or visual) can influence user behavior and/or trigger mistakes. Better 
	understanding of the effects of intended distractions will help clarify their role in adversarial models. 
	As part of the research effort described in this paper, we administered a range of naturally 
	occurring -- yet unexpected -- sounds while study participants attempted to perform a security-critical task. 
	We found that, although these auditory cues lowered participants' failure rates, they had no discernable 
	effect on their task completion times. To this end, we overview some relevant 
	literature that explains these somewhat counter-intuitive findings.

	Conducting a thorough and meaningful study on user errors requires a large number of participants, 
	since errors are typically infrequent and should not be instigated more than once per subject. 
	To reduce the effort of running numerous subjects, we developed a novel experimental setup 
	that was fully automated and unattended. We discuss our experience with this setup and 
	highlight the pros and cons of generalizing its usage.

	\end{abstract}

	\section{Introduction}
	\label{sec:intro}
	Our world is a noisy and distracting place, where truly quiet or sterile environments are rare. 
	Most people are accustomed to some degree of auditory and visual distraction in their daily lives. 
	However, they may be influenced in an unexpected manner by sudden distractions, 
	especially if they occur during performance of a task that demands concentration. 

	Meanwhile, modern technology allows -- and sometimes requires -- people to engage in security-critical 
	tasks in public settings, while being subjected to various degrees and types of sensory input. As personal 
	wireless devices (mainly smartphones) become more ubiquitous, the average person grows more reliant 
	on them for the performance of security tasks, such as entering a PIN, Bluetooth pairing or verifying 
	transaction amounts. For example, in online fund transfers, one has to compare the displayed amount 
	and currency to the intended amount and currency \cite{goodrich_using_2009}. In device pairing, 
	one needs to compare items (such as numbers, text, pictures, or sounds), or perform some physical 
	task over an ``out of band'' (OOB) channel  \cite{kobsa_serial_2009}. 

	All these tasks require some form of human involvement, which represents the weakest link and determines 
	overall security \cite{goodrich_using_2009, kobsa_serial_2009, nithyanand_groupthink:_2010, 
	kainda_usability_2009, kainda_two_2010, kobsa_can_2013, paul_field_2011}. This motivates extensive usability 
	studies to assess human ability to routinely complete 
	security tasks that still provide an acceptable level of security. There has been a lot of research on this topic 
	\cite{nithyanand_groupthink:_2010, kainda_usability_2009}, but very little work only that investigates user errors
	and maliciously induced user errors.
	One major reason for the dearth of prior work in this area is the difficulty of conducting traditional user 
	experiments. Since human errors  in such cases are relatively rare, it would take many trials with many 
	subjects to obtain statistically reliable information about the failure rate, and to determine whether the 
	difference in rates between two methods is statistically significant.\footnote{See Appendix.} The problem is exacerbated by the 
	fact that more than one method needs to be tested, while at the same time, only one attempt should be 
	made per study participant to trigger a mistake (since subjects may otherwise become alerted to such 
	attempts, consciously or subconsciously). For all these reasons, the total number of subjects needed 
	for an experiment to study user mistakes in security-related tasks can quickly grow into the hundreds.   

	To mitigate the effort needed to conduct such large studies, we designed a setup for an entirely unattended 
	experiment, wherein subjects receive recorded instructions from a life-size, video-projected, 
	rather than "live", experimenter.  As a first experiment in this environment, we decided to test the 
	error rate of subjects attempting to pair two Bluetooth devices in the presence of unexpected audio stimuli. 
	We tested $147$ subjects in this environment with no experimenter involvement. Our original expectation 
	was that unexpected audio interference would have a negative impact 
	on the completion of security-critical tasks. However, surprisingly, it turned out that noise actually had a facilitatory effect. 

	\noindent {\em{Organization:}} 
	Section \ref{sec:related} describes related work. 
	Then, Section \ref{sec:experiment} presents the design and setup of our experiments. 
	It is followed by Section \ref{sec:results} which presents experimental results. 
	Next, Sections \ref{sec:lessons} and  \ref{sec:discussion} summarize lessons learned
	from this experience and discuss conclusions, respectively.
	Section \ref{sec:limits} acknowledges certain limitations of our approach.
	Then, Section \ref{sec:ethics} addresses ethical considerations 
	and Section \ref{sec:future} concludes the paper with the discussion of future work.
	Appendix A provides a brief overview of statistical methods used and Appendix B
	demonstrates the unattended experiment setup.

	\section{Related Work}
	\label{sec:related}
	This section overviews related work in three areas: (1) automated experiments, (2) human-assisted 
	security methods, and (3) effects of noise on human task performance.

	\subsection{Automated Experiments}
	To the best of our knowledge, there has been no prior usability study utilizing a fully automated 
	(unattended) experimental setup with a video-projected experimenter. However, there have been 
	precedents with virtually attended experiments and unattended online surveys, in many fields, 
	most notably, psychology. There is a sizable body of work supporting validity and precision of 
	such unattended online experiments, as compared to more traditional attended experiments in a lab setting. 
	In particular, Ollesch et al.\ \cite{ollesch_physical_2006} found no significant difference in psychometric data collected from an 
	attended experiment in a lab setting and its online, virtually attended counterpart. This is further supported by  
	Riva et al.\ \cite{riva_use_2003} in the comparison of data collected from unattended online and attended offline questionnaires. 
	Guidelines for creating the best possible draw from the intended population base are provided by Birnbaum
	\cite{birnbaum_human_2004}. Finally, Lazem and Gracanin \cite{lazem_social_2010} replicated two classical 
	social psychology experiments, where the experimenter and the three participants were represented as avatars 
	in Second Life instead of being physically co-present. The outcomes were very similar. 

	However, all aforementioned studies assume that an unattended or virtually attended experiment 
	occurs online. There appears to be no prior work involving an offline, unattended  experimental 
	setup.

	\subsection{User Studies of Security Protocols}
	Both security and usability experts have extensively studied secure device 
	pairing. In this setting, wireless devices have no prior knowledge of one another and, hence, there is no 
	pre-existing security context. This is further complicated by the inability to rely on either a common 
	Public Key Infrastructure (PKI) or a mutually Trusted Third Party (TTP).  This accentuates threats of 
	man-in-the-middle (MitM) or "evil twin'' attacks during the pairing protocol. Consequently, involvement 
	of a human user has been proposed, in order to verify (over a low-bandwidth OOB channel) message 
	integrity of the protocol that transpires over the normal channel. 

	For example, Short Authentication String (SAS) protocols require a user to compare two short 
	strings, of about 20 bits each \cite{cryptoeprint:2005:424}. Since accurate task completion was 
	found to be relatively difficult for human users, alternative protocols were developed.

	The first usability study of pairing techniques was carried out by Uzun et al.\ \cite{Uzun_2007}. That study
	determined that the most accurate
	way to compare a pair of SAS was the "compare and confirm" method, wherein the user would be presented
	an SAS by both of the machines they are trying to pair, and would be to asked to confirm whether or not the two SASs match.

	Goodrich et al.\ \cite{goodrich_using_2009} introduced an authentication technique that utilized a ``Mad-Lib'' type structure, 
	where participating devices, based on the protocol outcome, compose a nonsensical phrase 
	out of several short English words. The human user is then tasked with determining whether 
	the two devices came up with matching phrases. This technique was found to be easier to 
	complete by non-specialist users. 

	Kobsa et al.\ \cite{kobsa_serial_2009} reported on a comprehensive comparative usability study of eleven major secure 
	device pairing methods, measuring task performance times, task completion rates, perceived security 
	and perceived usability. The main outcome was the grouping of the investigated methods into three 
	clusters, following a principal components analysis.

	Kainda et al.\ \cite{kainda_usability_2009} examined usability of device pairing in a group setting, where up to 6 users tried to 
	connect their devices to one another, and found that group effort decreased the expected rate of 
	security and non-security failures. Although, an inherent ``insecurity of conformity'' was also identified, 
	wherein users would deliberately lie about an observed string in order to ``fit in'' with the majority 
	opinion of a group.

	Nithyanand et al.\ \cite{nithyanand_groupthink:_2010} also examined the pairing of multiple devices in a group 
	setting with groups of 4 or 6 members. 
	They found that groups are nearly  immune to insertion attacks, where an adversary will pretend to be a member
	of the group, and thus change the expected SAS for all members. They also found that groups are
	particularly vulnerable to a modified man-in-the-middle attack where a single member of the group is given false information,
	and instead of rejecting their incorrect SAS, they conform to the positive result the rest of the group reports.

	Gallego et al.\ \cite{sadeghi_exploring_2013} found that performance on out-of-band tasks in secure device pairing could be 
	improved through the addition of a score metric on the user's performance, resulting in a 
	considerable reduction in both safe and fatal errors.

	There are numerous other results in the area of secure wireless device pairing. However, most of them 
	focused on newly proposed techniques rather than on comparative usability. 

	\subsection{Impact of Noise on Task Performance}

	\thispagestyle{plain}
	\pagestyle{plain}

	In the field of psychology, there are conflicting results with respect to the influence of noise on human 
	task performance. Some experiments claim a positive effect \cite{hockey_effect_1970,
	omalley_noise-induced_1971,baker_effects_1993}, while others report exactly the 
	opposite \cite{childs_effects_1972, benignus_effect_1975}. Several explanations of this phenomenon 
	have been proposed. Initially, it was 
	thought that the type of noise used was the primary factor that would cause either an inhibitory or 
	facilitatory effect. Hockey \cite{hockey_effect_1970} demonstrated though that this is not the case, as studies utilizing a 
	diverse range of audio conditions have reported both possible outcomes.

	The subsequent explanation was that task complexity might have something to with the effect of 
	noise upon the completion of that task. Hockey \cite{hockey_effect_1970} and Benignus et al.\ \cite{benignus_effect_1975} have 
	shown that such a causal 
	relationship may exist. Task complexity is defined by the task's event rate (i.e., how many elements of the 
	task are received within a given period of time), and by the number of different sources from 
	which these task elements are received. Tasks that have a low event rate and a low number 
	of sources are more likely to be facilitated than impaired by the introduction of noise. In contrast, 
	tasks with high event rates and many sources are more likely to be impaired by noise. 
	This hypothesized relationship views noise as a general stimulant that heightens general sensory arousal: 
	if a subject is at a very low level of arousal before the introduction of the noise, it can help sharpen their 
	focus and improve task performance \cite{olmedo_maintenance_1977, koelega_noise_1986}. 
	However, for a subject who is already at a high level of 
	sensory arousal, the added stressor can overload them and introduce errors in 
	task completion \cite{childs_effects_1972, harris_stress_1960}.

	Furthermore, O'Malley and Poplawsky \cite{omalley_noise-induced_1971} showed that noise can affect behavioral selectivity. 
	This means that while noise may not have a consistent positive or negative impact on task completion 
	in all cases, noise may consistently have a negative effect on tasks that require the subject to detect 
	signals in their periphery, and noise may have a consistent positive effect on task completion when 
	the subject has to focus on signals coming from the center of their field of attention. This suggests 
	that, regardless of task complexity, the addition of noise may narrow a subject's area of attention.

	\section{Experiment}
	\label{sec:experiment}
	This section describes our experimental setup, procedures and subject parameters.

	\begin{figure*}[htbp]
	\fbox{\centering
	\includegraphics[height=2.4in,width=2.1\columnwidth]{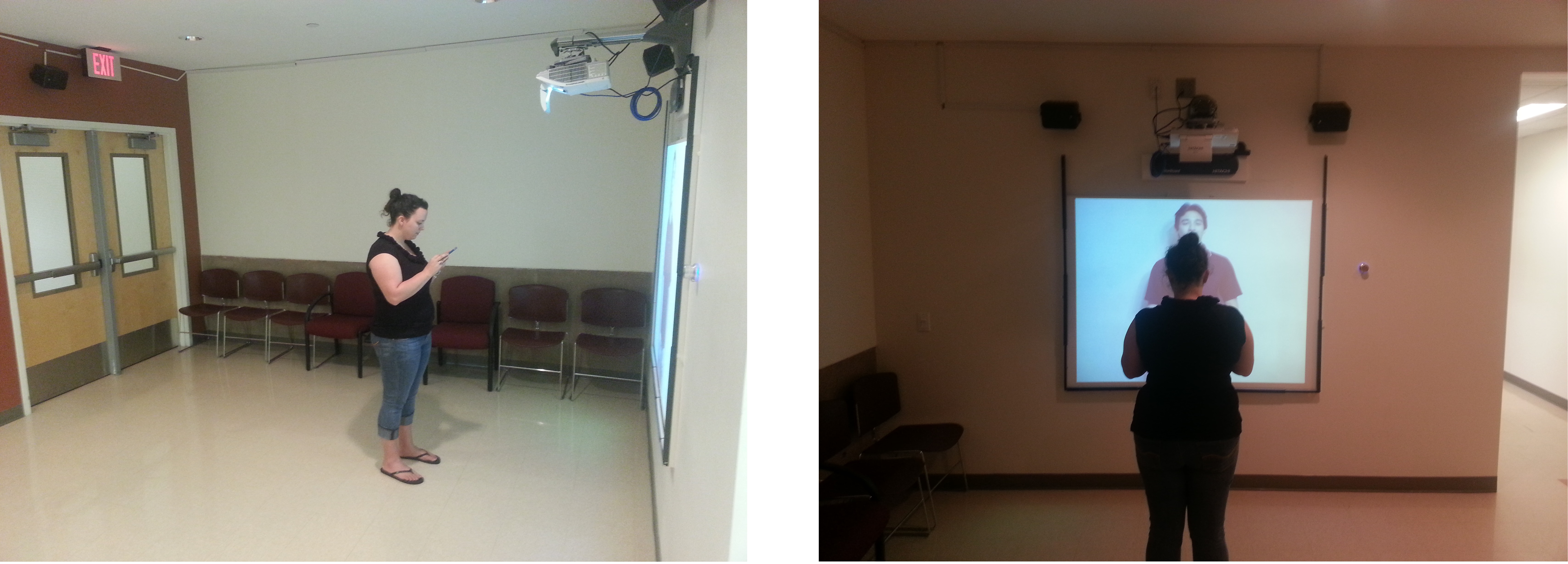}}
	\caption{{\small Experimental Setup: (a) Side view (speakers over the door),  (b) Front view} }
	\label{fig:setup}
	\end{figure*} 

	\subsection{Apparatus}
	The setting of our study was carefully designed to facilitate fully automated experiments 
	with a variety of sensory inputs. The installation is situated in a low-traffic public space (a wide 
	corridor corner nook) at the top floor of a large academic building on a university campus. 

	Figure \ref{fig:setup}(a) shows the experimental location from the side, and 
	Figure \ref{fig:setup}(b) shows our setup from the subject's perspective (front view).
	It includes a large touch-sensitive 
	Smartboard with a short-throw projector, a webcam, and two pairs of speakers 
	(one in front and one behind the intended subject position), a motion detector, as well as 
	controllable lights and electricity outlets. The Smartboard is an interactive whiteboard 
	(see {\url{smarttech.com}}) that gathers input via user's touch on its surface. 
	As such, it acts as both the display and the input device. 

	Instead of a human experimenter actively curating the environment 
	and interacting with the subjects, we used a life-size video/audio recording of an experimenter as a proxy. 
	This proxy is the subject's main source of information about the experiment. In particular, the proxy 
	starts by reading a script explaining the flow of the experiment. This is shown in Figure \ref{fig:proxy}.

	This setup allows for a fully unattended experiment. The only (and strictly off-line) involvement of an 
	experimenter amounted to infrequent re-calibration of sound effects and repair of some components 
	that suffered (minor) damage throughout the study.
	 
	\subsection{Procedures} \label{proc}
	The goal of the experiment was to measure user errors when attempting to pair two wireless devices 
	via Bluetooth, while being exposed to potentially distracting and possibly ``malicious'' sound effects. 
	To make the ``attack'' less noticeable, we use four sounds that can be encountered in real life, both in open 
	and enclosed public spaces: (1) a baby crying, (2) a hammer striking a wall, (3) helicopter rotors spinning, and 
	(4) a circular saw cutting wood.  Reasons for selecting these four specific sounds as audio stimuli are 
	discussed in Section \ref{divstim} below.

	All sounds were played at normal volume from a set of speakers situated behind the subject. Specific 
	volumes of the four sounds (measured at a typical subject's position) were as follows: 
	\begin{compactitem}
	\item Baby: 67 dB
	\item Helicopter: 79 dB 
	\item Hammer: 80 dB 
	\item Saw: 78 dB
	\end{compactitem}
	Even the highest of these four volumes (80 dB) is well within the {\em safe range}, as defined by the US 
	Occupational Safety \& Health Administration (OSHA) guidelines.\footnote{
	OSHA requires all employers to implement a Hearing Conservation Program where workers are 
	exposed to a time-weighted average noise level of 85 dB or higher over an 8 hour work shift. Our
	noise levels were clearly lower. See: {\url{https://www.osha.gov/SLTC/noisehearingconservation/}}}

	To begin the experiment, the subject approaches the Smartboard and presses a large wall-mounted 
	button to the right. Although a motion-activated start is also possible, we decided to minimize 
	any disturbance for uninvolved passers-by. Next, the Smartboard plays a short video recording of our proxy 
	experimenter, who explains that the subject will be preforming a task on their own phone, namely, 
	connecting it via Bluetooth to a nearby device. The latter is actually an iMac desktop in the office behind the 
	Smartboard; it is not visible to the subject, as shown in Figure \ref{fig:review}. 
	The subject is promised a reward for the successful completion 
	of the experiment, in the form of a $\$5$ Amazon coupon. The subject is also briefly informed that the 
	task of pairing two Bluetooth devices involves comparing two 6-digit numbers and confirming 
	whether they match, as shown in Figure \ref{fig:confirm}.

	At this point, the subject has a time window of 2 minutes to correctly pair the devices. Otherwise, 
	a failure message is read out and displayed. While the subject is in the process of pairing, 
	one of five events occurs: either silence is maintained throughout the experiment, or one of 
	the aforementioned four sounds is played from the speakers located on the ceiling behind the subject.

	A subject who fails the first time and wishes to make another attempt at pairing, is given the 
	opportunity to re-try the experiment in another two-minute window. 
	If pairing completes successfully, a message to that effect is displayed.  At the end, a subject 
	is asked to enter an email address using a virtual keyboard displayed on the touch-sensitive Smartboard 
	(see Figure \ref{fig:email}), thus allowing us to email the promised Amazon coupon as a participation reward. 

	Each subject encountered only one condition. Presenting subjects with two conditions would have biased 
	their performance in the second condition, since, at that point, they would already know what to do and what 
	might happen. Since subject observables (errors) are influenced by various individual characteristics, 
	random subject selection ensures that any variation between sample and population observables is only a matter of chance.

	After successful completion of the experiment, if the same subject attempts to repeat the same experiment 
	with the same personal device, their data is automatically flagged and later discarded. Multiple participation 
	of the same subject with different personal devices is identified (and discarded) by visual inspection of 
	video recordings. The experimental setup maintains a detailed log of all system events that can later 
	be analyzed to measure outcomes, such as the number of re-trials, task success rates, and task completion times,
	as well as a video recording of the entire encounter, as shown in Figure \ref{fig:review}.

	\subsection{Hypotheses} \label{hypo}
	Our initial hypotheses were that introducing noise while an unsuspecting subject attempts to pair 
	two Bluetooth devices will have no effect:
	\begin{compactitem}
	\item[H1] We will observe the same error rate 

	and

	\item[H2] The pairing process will take the same amount of time to complete successfully, 
	\end{compactitem}
	as in the same setting without any noise interference.

	\subsection{Subjects}
	In prior studies on usability of pairing protocols with a human in the loop \cite{goodrich_using_2009}, 
	\cite{nithyanand_groupthink:_2010}, \cite{kainda_usability_2009}, it was discovered 
	that a subject population of 20-25 per condition being tested was an acceptable size for 
	obtaining statistically significant\footnote{See Appendix for the definition of
	statistical significance.} findings. Since our planned experiment has one condition for 
	each of the four sound effects as well as one control condition (with no sound), collecting any 
	meaningful amount of data would require well over one hundred iterations of the experiment. 

	To recruit subjects, we posted signs around the entrance and inside the lobby of a large campus 
	building, which directed people to the experimental setup and mentioned the reward for participation. 
	Posters explicitly described that subjects were sought for a brief "Usability Study" and did not in any 
	way mention the security-critical nature of the task to be performed, or the possibility of 
	any noise interference. The general area of campus where the experiments were conducted houses 
	Computer Science and Engineering departments.

	Of the total 147 subjects, there were 102 males and 45 females. Most of them (139 out of 147) 
	appeared to be college-aged (18-24 years), while 8 seemed to belong to a somewhat older group (30+ years). 
	This demographic breakdown is influenced by the location of the experiment and by the recruitment form. 
	Since we solicited participants passively and since our recruitment posters were
	located in the "technical'' part of a large university campus, it is not surprising that the overwhelming majority of 
	participants were of college age with the majority being male.

	\section{Results}
	\label{sec:results}
	We now discuss the results of the study, starting with data cleaning and 
	proceeding to task completion results. Statistical tools used  in data analysis
	are described in the Appendix.

	\subsection{Data Cleaning}
	\label{subsec:cleaning}
	Subject data was discarded in three cases. First, we removed the instances where participants arrived either in 
	pairs or larger groups. Their data were eliminated since it might have been skewed due to social facilitation. 
	It has been shown that being 
	under observation of others can have a positive impact on subjects performing tasks of low 
	levels of complexity \cite{aiello_social_2001}. Second, a few participants arrived with 
	old-style flip phones. Such older phones were technically unable 
	to establish a Bluetooth connection with our client. 

	All in all, 29 pairs or groups of subjects had to be discarded, as well as 10 others who attempted to use flip phones. 
	We could not discern any obvious visual or auditory impairment in any subject that would be a detriment to the 
	experiment. We later visually checked all experiments for subjects with such impairments and none were identified.

	Finally, in discussing results below, we differentiate between pairings and failures that 
	occurred in the first two-minute trial window, and those that occurred across all attempts. 
	While the differences are not large, the former results may capture those subjects 
	better that have already had some pairing experience in the past.

	\subsection{Task Completion Rate}
	Table \ref{tab:FR1} shows the numbers of subjects whose first attempt at pairing resulted in a 
	success and failure, respectively, plus the failure rate for the control condition and each stimulus condition. 

	\begin{table}[!htb]
	\centering 
	\caption{\small Subject failure rate, first attempt only}
	\label{tab:FR1}
	{\setlength{\extrarowheight}{10pt}
	\begin{tabular}{||c|c|c|c||}
	\hline \cline{1-4}
	Stimulus &  \#Successful & \#Unsuccessful  & Failure \vspace*{-0.3cm} \\
					&  Subjects & Subjects & Rate
	\\ \hline \cline{1-4}
	None (control)	& 27 	& 13 	& 0.34  
	\\  \hline
	Baby &	        23 &	1	& 0.04
	\\ \hline
	Hammering &	33 &	3	& 0.08
	\\ \hline
	Helicopter &	24 &	1	& 0.04
	\\ \hline
	Saw &           	20 & 2	 & 0.09
	\\ \hline
	{\bf Total} &           	127 & 20	& 0.14
	\\ \hline \cline{1-4}
	\end{tabular}}
	\end{table}

	Table \ref{tab:BT} shows the parameters for the Barnard's exact test applied pairwise to the 
	subject failure rate of the control condition and each stimulus. It shows that differences between 
	failure rates are statistically significant $(p<0.05)$ with respect to all four stimuli. This also holds 
	if one applies a conservative Bonferroni correction to account for four pairwise comparison (see the Appendix), 
	which leads us to reject hypothesis H1 in Section \ref{hypo}, since the failure rate significantly decreases with the 
	introduction of noise. Section \ref{sec:discussion} discusses this further.

	\begin{table}[!h]
	\centering 
	\caption{Barnard's Exact Test on subject failure rates of control \& stimuli}
	\label{tab:BT}
	{\setlength{\extrarowheight}{10pt}
	\begin{tabular}{||c|c|c|c|c|c||}
	\hline \cline{1-6}
	Stimulus & Total & Failure & Wald & Nuisance & {\large \it $p$} 
	\vspace*{-0.3cm} \\
				  &  Pairings & Rate &  Statistic & Parameter & 
	\\ \hline
	None(control) & 40 & 0.34& --&--&--
	\\ \hline
	Baby & 24  & 0.04 & 2.65 & 0.95 & 0.03
	\\ \hline
	Hammering & 36 & 0.08 & 2.58 & 0.91 & 0.01
	\\ \hline
	Helicopter & 25 & 0.04 & 2.71 & 0.89 & 0.01
	\\ \hline
	Saw &           	22 & 0.09	& 2.05 & 0.84 &0.03
	\\ \hline \cline{1-6}
	\end{tabular}}
	\end{table}

Table \ref{tab:Odds} shows odds ratios and 95\% confidence interval for each stimulua compared to the control condition. 
Interestingly, under this analysis, confidence interval of the Saw condition includes a possible odds ratio of $1.0$. This 
implies that -- under this method of analysis -- it is not statistically significant at the 95\% level. Confidence 
intervals for other 3 stimuli reinforce the claim of statistical significance at the 95\% level, as established by 
Barnard's exact test.  

	\begin{table}[!htb]
	\centering 
	\caption{\small Odds Ratio and 95\% Confidence Intervals on Subject Failure Rates of Control and Stimuli}
	\label{tab:Odds}
	{\setlength{\extrarowheight}{10pt}
	\begin{tabular}{||c|c|c||}
	\hline \cline{1-3}
	Stimulus &  Odds Ratio  &  95\% Confidence Interval
	\vspace*{-0.3cm} \\
		&	wrt control & wrt control
	\\ \hline \cline{1-3}
	None (control)	& -	& -- 
	\\  \hline
	Baby &	        0.09 &	0.01 - 0.74
	\\ \hline
	Hammering &	 0.18 & 0.04 - 0.73
	\\ \hline
	Helicopter &	0.09 & 0.01 - 0.71
	\\ \hline
	Saw &           	0.20 & 0.04 - 1.02	 
	\\ \hline \cline{1-3}
	\end{tabular}}
	\end{table}

	Table \ref{tab:FR2} shows the total number of pairing attempts that ended in success 
	(as well as those that failed)  across {\em all pairing trials}, and the failure rate for each
	stimulus (and control) condition.  
	We note that not every subject who initially failed chose to re-try. However, 
	every subject who re-tried was successful the second time.

	\begin{table}[h!]
	\centering
	\caption{Failure rate by stimulus across all attempts}
	\label{tab:FR2}
	{\setlength{\extrarowheight}{10pt}
	\begin{tabular}{||c|c|c|c||}
	\hline \cline{1-4}
	Stimulus &  \#Successful & \#Failed  & Failure \vspace*{-0.3cm} \\
					&  Pairings & Pairings & Rate 
	\\ \hline \cline{1-4}
	None (control)	& 28	& 13	& 0.32
	\\ \hline
	Baby		   & 24	& 1	& 0.04
	\\ \hline
	Hammering & 34	& 3	& 0.08
	\\ \hline
	Helicopter	 & 24	 	& 1	& 0.04
	\\ \hline
	Saw	 & 20	 	& 2	& 0.09
	\\ \hline
	{\bf Total}	& 130	& 20	 & 0.13
	\\ \hline \cline{1-4}
	\end{tabular}}
	\end{table}

	We also partitioned subject failure rates by gender. While Table \ref{tab:gender}
	seems to indicate that female subjects were substantially less likely to fail on the initial 
	attempt than their male counterparts, performing Barnard's exact test on the 
	subject failure rates of men and women revealed that the perceived 
	difference between them is not statistically significant; Wald statistic = 2.32, nuisance parameter = 0.98, $p = 0.14$.

	\begin{table}[!htb]
	\centering 
	\caption{Subject failure rate by gender, first attempt only}
	\label{tab:gender}
	{\setlength{\extrarowheight}{10pt}
	\begin{tabular}{||c|c|c|c||}
	\hline \cline{1-4}
	Gender &  \#Successful & \#Unsuccessful  & Failure \vspace*{-0.3cm} \\
				 &  Subjects       & Subjects              & Rate
	\\ \hline \cline{1-4}
	Male	&  86	& 16 	& 0.16
	\\  \hline
	Female &	       41 &	4	& 0.09
	\\ \hline \cline{1-4}
	\end{tabular}}
	\end{table}

	\subsection{Task Completion Times}
	Table \ref{tab:times} shows average completion times in successful trials for subjects under each stimulus. 
	After applying a conservative Bonferroni correction to account for four pairwise comparisons, there is
	no statistically significant difference in completion times between the control condition and each stimulus. 

	\begin{table}[ht!]
	\centering 
	\caption{\small Avg times (sec) for successful pairing}
	\label{tab:times}
	{\setlength{\extrarowheight}{10pt}
	\begin{tabular}{||c|c|c|c|c|c||}
	\hline \cline{1-6}
	Stimulus & Mean  & Standard & DF wrt & t-value  &  {\large \it $p$}
	\vspace*{-0.3cm} \\
	 & Time & Deviation &  control & wrt control & 
	\\ \hline \cline{1-6}
	None & 34.41 & 13.78 & -- & -- & -- 
	\\ \hline
	Baby & 31.13 & 10.06 & 63 & 0.97 & 0.35
	\\ \hline
	Hammering & 28.82 & 9.76 &74 & 1.84 & 0.07
	\\ \hline
	Helicopter & 31.33 & 13.13 & 63 & 0.81 & 0.39 
	\\ \hline
	Saw & 38.45& 17.15 & 60 & 0.90 & 0.38 
	\\ \hline \cline{1-6}
	\end{tabular}}
	\end{table}

	Table \ref{tab:Cohen} shows Cohen's $d$ and its 95\% confidence interval, for subject 
	completion times under each of the stimuli when compared to the control condition. 
	The effect sizes of the stimuli are not statistically significant from 0 since each of the confidence intervals contains 0.

	\begin{table}[!htb]
	\centering 
	\caption{\small Cohen's $d$ and 95\% Confidence Intervals on Subject Completion Times Between Control and Stimuli}
	\label{tab:Cohen}
	{\setlength{\extrarowheight}{10pt}
	\begin{tabular}{||c|c|c||}
	\hline \cline{1-3}
	Stimulus &  Cohen's d  &  95\% Confidence Interval
	\vspace*{-0.3cm} \\
		&	wrt control & wrt control
	\\ \hline \cline{1-3}
	None (control)	& -	& -- 
	\\  \hline
	Baby &	        0.27 &	-4.00 to 4.29
	\\ \hline
	Hammering &	 0.47 & -3.80 - 3.66
	\\ \hline
	Helicopter &	0.23& -4.04 - 5.48
	\\ \hline
	Saw &           	-0.27 & -4.54 - 6.89	 
	\\ \hline \cline{1-3}
	\end{tabular}}
	\end{table}

	As with subject failure rates, we also examined subjects' completion times for successful pairing attempts by gender.
	The results are displayed in Table VIII. A pairwise t-test shows that the observed differences are not statistically significant (t(148) = 1.23, p = 0.22). 

	\begin{table}[ht!]
	\centering 
	\caption{\small Avg times (sec) for successful pairing by gender}
	\label{tab:gendertimes}
	{\setlength{\extrarowheight}{10pt}
	\begin{tabular}{||c|c|c||}
	\hline \cline{1-3}
	Gender & Mean  & Standard 
	\vspace*{-0.3cm} \\
	  & Time    & Deviation 
	\\ \hline \cline{1-3}
	Male & 30.63 & 10.92
	\\ \hline
	Female & 33.23 & 13.85
	\\ \hline \cline{1-3}
	\end{tabular}}
	\end{table}

	\section{Lessons Learned}
	\label{sec:lessons}
	As mentioned above, some subjects participated in the experiment in pairs. We had not explicitly forbidden 
	this since doing so in an unattended setting would be impossible. We ignored the data of such 
	participant pairs, see Section \ref{subsec:cleaning}. A few subjects also tried the experiment more than once on 
	different Bluetooth devices (presumably to earn the participation reward multiple times), and we 
	had to visually identify and discard their data.

	Furthermore, a few subjects did not understand how to pair two devices using Bluetooth, or 
	were unsure what they were supposed to do in general. This illustrates one drawback with our 
	experiment design - there was no option to replay the instructions, nor was there a set of more 
	detailed instructions for participants who were unfamiliar with the Bluetooth functionality of their 
	devices. Since our experiment was unattended, there was no way to tell the cause of task failure 
	in real time, or to help the subject if needed, until the recording of the subject's 
	trial was viewed.

	Interestingly, quite a few subjects had trouble following the instructions of the proxy  
	to enter their email address on a virtual keyboard that was projected onto the touch-sensitive 
	Smartboard. Up to this point, the Smartboard had only served as a (completely passive) projection 
	wall, and subjects may have been surprised that it could also be used as an input device.

	Our experiments were conducted during the academic year while the term was in session. 
	During that time, even though the area where recruitment posters were placed experienced heavy
	foot traffic, our prominently placed signs did not attract as many subjects in a short period of time (1-2 weeks)
	as we had expected. Therefore, in order to gain a sufficient number of subjects,  
	the experiment lasted about 6 weeks with some short breaks when recruitment posters were removed. 
	These short breaks actually proved useful, since we believe that periodic 
	appearance and re-appearance of recruitment posters motivated additional participants.

	In retrospect, video surveillance of the experimental setup proved invaluable, both for actual security 
	purposes and for being able to later correct experimental lapses.

	\section{Discussion of Observed Effects} 
	\label{sec:discussion}
	The introduction of several types of peripheral audio noise did not appear to interfere with the completion
	of the task of Bluetooth pairing. In fact, collected data shows a significant decrease in failure rates for every 
	stimulus, with no statistically significant difference in the failure rate between different noise stimuli. 
	In cases of baby crying and helicopter's rotors spinning, there was only a single 
	failure across 25 attempts, as opposed to the control case, where every third attempt resulted in 
	failure, as shown in Table \ref{tab:FR1}.

	This result, while initially unexpected, is actually consistent with the Brain Arousal Model of 
	\cite{soderlund_positive_2008} as well as the results of \cite{benignus_effect_1975,hockey_effect_1970,omalley_noise-induced_1971}. 
	Our experiment has a single, centrally-located source and a low event rate, so it would be reasonable to 
	expect noise to have a positive effect on successful task completion rates \cite{benignus_effect_1975, hockey_effect_1970}. 
	Audio signals in our study came from the center of participants' area of attention and not the periphery, 
	i.e., the Smartboard and participants' smartphones were in front of them. This suggests that noise caused 
	participants to narrow their focus of attention \cite{omalley_noise-induced_1971}, which might be conducive to 
	better task performance.
	Nevertheless, our study is novel in the context of security-critical tasks, since, by its very nature, the process 
	of Bluetooth device pairing is significantly shorter in duration than the attention-intensive vigilance tasks 
	discussed in related literature.

	\section{Limitations}
	\label{sec:limits}
	We readily acknowledge that the study described in this paper, although the first of its kind, has certain 
	shortcomings and limitations, detailed below.

	\subsection{Subjects}
	We experimented with a narrow subject group, dominated by young and tech-savvy
	college students. This is a direct consequence of the specific campus location of our unattended setup. 
	Replicating it in a non-academic setting (e.g., an office building) would be possible and
	useful. However, passive recruitment of a really diverse group of participants is only possible in a truly
	public space with a high volume of traffic, e.g., a stadium, a shopping mall, a movie theater or a concert hall. 
	On the other hand,  placing our unattended experiment setup in any of such settings would be extremely challenging. First, our setup involved specialized and expensive equipment whose security would be difficult
	to ensure in a very public space. Second, high-traffic public spaces tend to be have lots of background or ambient
	noise which would interfere with the stimuli in our experiments.

	As already mentioned, the nature of our location also had a skewed impact on the gender breakdown of our subjects.
	Since the experiment was set up in the Computer Science and Engineering section of a large university campus, 
	the majority of the passers-by were male. Because of this, we were unable to collect sufficient data in a realistic time 
	frame to examine the effects of each individual stimulus on subjects of each gender. 

	One potential problem with our subjects is that young people are in general more sensitive to 
	noise than older adults \cite{Brant_1990}. It is quite conceivable that older and/or 
	technologically non-adept people\footnote{Since people who are new to, or unfamiliar with, 
	a specific technological task would naturally be more nervous or tense when performing it.} 
	would react differently to our noise stimuli. 

	Finally, recall that our experimental setup required the subject to interact with both visual and audio queues from the proxy 
	experimenter and the environment. Because of this, an ideal subject would have no substantial hearing or
	visual impairment. However, due to the unattended nature of our experiment, we could not proactively
	rule out such subjects (e.g., by specifying restrictions in the recruitment posters) 
	without  giving away the nature of our experiment. Doing so would have created an initial
	expectation for subjects who fit our criteria, which could adversely influence accuracy of collected data. 
	Therefore, during later review of each video-recorded experiment, we had to verify that there were no participants
	with obvious visual and/or hearing impairments.

	\subsection{Diversity of Stimuli} \label{divstim}
	We experimented with four stimuli through a subjective process of elimination, with the intention of getting as 
	many diverse noise types as we could rigorously test, that were annoying to the listener in varying degrees. 
	With respect to diversity, we classified sounds in three ways: 
	\begin{compactenum}
	  \item Continuous or discrete 
	  \item Regular or irregular 
	  \item Human-generated or synthetic/mechanical
	\end{compactenum}
	One reason for settling on such a small number of stimuli was due to the combination of (1) the 
	location of the experiment, and (2) placement of study recruitment posters.
	Although posters were placed in a high-traffic zone, outside the building where the experiments
	took place, the same people (mostly students) tend to walk by every day due to the regularity of campus life, e.g., 
	classes begin and end at the same time and at the same place. Consequently, although we were able to 
	attract 147 subjects, the rate of participation decreased markedly over time and ceased completely after 6 weeks.
	As it turned out, 147 was just enough for four stimuli as well as the control condition.
	An additional stimulus would have needed around 25 new subjects; that proved impossible under the conditions
	of our study.\footnote{Of course,  recruitment posters could have distributed better around campus. However,
	experience shows that attracting participants from farther afield is harder than from nearby locations, especially
	given the relatively meager participation reward.}

	Despite this constraint, we selected the four stimuli to be as diverse as possible: 
	\begin{compactitem}
	\item Baby crying was a continuous, irregular, human-generated sound
	\item Helicopter rotors was a continuous, regular, mechanical sound
	\item Hammering was a discrete, regular mechanical sound, and
	\item Circular saw was a continuous, irregular mechanical sound 
	\end{compactitem}
	The most obvious discrete, human-generated stimulus -- talking -- was intentionally 
	omitted, since it would have likely caused confusion between the experiment instructions and 
	the stimulus.

	\subsection{Insufficiently Security-Critical Task}
	We suspect that most participants were unaware, ahead of time, of the purpose and
	details of our experiment. However, during the experiment they clearly understood 
	that the task at hand was Bluetooth-based pairing of their smartphone with some other (our) device.
	Consequently, from the participant's perspective, this task was unlikely to be perceived 
	as being truly security-critical; the device the subjects were asked to connect to was obviously a prop, not a 
	device the subject owned.

	In the same vein, device pairing is neither as security-critical nor as pervasive 
	(or frequent) as other tasks, such as password or PIN entry for the purpose of 
	Internet access or PIN entry into an Automated Teller Machine (ATM). 
	However, experimenting with these more natural tasks is significantly more 
	difficult.

	\subsection{Synthetic Environment}
	Our unattended experiment setup is clearly very synthetic, for several reasons: First, 
	it is normally very quiet, unlike many (perhaps most) common everyday settings.
	Second, it is located indoors with no exposure to daylight, no air movement and
	no temperature fluctuations. Third, the setup (as shown in Figure \ref{fig:setup})
	involves equipment that an average participant never or rarely encounters in 
	the real world, in particular, a touch-sensitive Smartboard used as a means of 
	both input and output, and a unusual-looking companion projector. 

	\subsection{Ideal Setting} \label{ideal}
	Based on the above discussion, it is easy to see that the ideal setting for our experiment 
	would be one where:
	\begin{compactitem}
	\item Demographics of participants is widely varied
	\item Participants are completely unaware of the experiment, at least until it is over
	\item The environment is common/natural
	\item The task is truly security-critical
	\end{compactitem}
	One trivial example of such an ideal setting is a bank ATM 
	located in a well-trafficked public space, with the security-critical task being the PIN
	entry process. A modern ATM incorporates all features
	needed for our type of experiments: a keypad, a screen, a speaker (for visually impaired
	individuals), and a video camera. A similar setting is encountered in some automotive
	gas stations where the fuel pump includes a keypad (used for PIN and/or Zip code entry), 
	a screen and a speaker; video cameras are usually located overhead. Yet another example
	would be a setting with public Internet access terminals, commonly found in airports and hotels,
	where the security-critical task would be the log-in process to the Internet provider. 

	In theory, in any of the above examples, large numbers of diverse subjects can be seamlessly 
	gathered without any explicit recruitment, awareness of the experiment or reward for participation.
	However, it is easy to see that conducting experiments in these ideal settings would be
	physically, logistically and ethically problematic.

	\section{Ethical Considerations}
	\label{sec:ethics}
	Experiments described in this paper were fully authorized by the Institutional Review Board (IRB) of 
	our university, well ahead of the actual commencement of the study. The level of review was: 
	Exempt, Category II. 
	Further IRB-related details are available upon request.
	We note that no sensitive data was harvested during the experiments and minimal
	identifying information was retained. In particular:
	\begin{compactitem}
	\item  As part of Bluetooth device pairing, participants were not asked to select any 
	secret PINs or passwords. Instead, the 6-digit PIN was generated on the computer
	hidden from view and displayed on the Smartboard as well as their smartphone; they were 
	then asked to compare the two PINs and confirm that they were identical.
	\item The hidden computer (iMac) used for pairing was periodically flushed of all 
	collected device pairings.
	\item No names, addresses, phone numbers or other identifying information was collected
	from the participants.
	\item Although email addresses were solicited in order to deliver the participation reward,
	they were erased very soon thereafter.
	\item Video recordings of the experiments were (and still are) kept for study integrity 
	purposes. However, we plan is to erase them before IRB expiration time. 
	\end{compactitem}
	Finally, with regard to safety, we maintained noise levels of between $67$ and $80$ dB
	which is (especially for a very short duration, i.e., less than a minute) generally considered
	safe for people, as discussed earlier in Section \ref{proc}.

	\section{Future Work}
	\label{sec:future}
	As the ``human link'' in security-critical tasks becomes more popular in various settings, including
	those subject to accidental or adversarial sensory input, 
	a thorough evaluation of usability in the context of such tasks becomes imperative. 
	This work took the first step by studying the effects of unexpected audio noise on users
	performing wireless device pairing. 

	An interesting next step would be to conduct 
	a similar experiment with various visual stimuli, that is, to model an adversary who controls 
	some aspect of the visual environment of the experiment. The natural follow-on then would be
	to investigate the effects of mixed audio-visual stimuli.

	Our use of an unattended experiment led to several complications that we had not anticipated. 
	For example, we were often confronted with multiple subjects simultaneously taking part in the 
	experiment or advising one another on how to correctly complete the task at hand. A technical 
	solution to this problem would be an enclosed experimental area whose access is restricted to 
	entry by a single person only (e.g., through a controlled turnstile). Unfortunately, this would be in 
	violation of fire safety regulations. We therefore plan to explicitly instruct participants that the 
	experiment is intended to be conducted by a single subject at a time, and to verify and penalize 
	non-compliance, e.g., by denying reward to non-compliant subjects.

	Alternatively, instead of discarding these results, in future studies it may be worthwhile to 
	consider and compare the results of those collaborating subjects' trials in the context of all 
	trials with multiple participants. Such a comparison was beyond the scope of our initial experiment. 

	To proactively discourage multiple experiments by the same subject with different Bluetooth devices 
	we could explicitly advertise the fact that video recordings will be reviewed and subjects who 
	participate more than once will not receive a reward. However, this would accentuate the fact that 
	subjects are on camera, which could potentially influence performance.

	We feel that this experimental paradigm is valuable and deserves further evaluation. 
	One possible goal is to create a new standard whereby large experiments with 
	hundreds of subjects can be conducted without posing a prohibitive financial and/or logistical burden.

	\section*{Acknowledgments}
	\label{sec:ack}
         This research was supported by NSF grant CNS-0831526.
         We would like to thank all anonymous participants in our study.

	\bibliographystyle{ieeetr} 
	\bibliography{noise-sec}

	\appendix

	\section*{Appendix A: Statistical Tools}
	\label{sec:stats}
	This section overviews simple statistical tools utilized in the analysis of collected data. 

	\subsection{Statistical significance}
	\label{statsig}
	A statement of statistical significance is a statement of confidence that two sets of observations represent 
	samples taken from different populations. This claim is constructed by comparing the probability that two 
	sets of observations could have been taken from the same population distribution. The probability $p$ of the two samplings 
	being from the same population is then evaluated against a low-threshold, $\alpha$. If $p \leq \alpha$, the result is said 
	to be {\em statistically significant}, and one could confidently claim that the two samples represent two different populations. 
	We fix the low-threshold at $\alpha = 0.05$.

	\subsection{Barnard's Exact Test}
	\label{subsec:barnard}
	Barnard \cite{Barnard_1947} specifies a method for testing  independence of rows and columns in a 
	contingency table that takes into account nuisance parameters, which are auxiliary parameters 
	that may not be of immediate interest, but instead have a direct impact on the parameter(s) 
	being evaluated. Barnard's exact test seeks to maximize the impact of any such nuisance 
	parameters in maximizing $p$, the likelihood that the two rows in the contingency table are 
	samples taken from the same population. This means that the $p$ values found by this test 
	are worst-case estimates, and are more suitable for extreme cases than similar methods, 
	and are at least as powerful in non-extreme cases. We utilize Barnard's exact test in the
	examination of subject failure rates.

	\subsection{Paired t-tests}
	The t-test is a method for determining if the differences between two data sets are statistically 
	significant through the examination of their mean and standard deviation. For populations that can 
	be assumed to follow a normal distribution, these two parameters are sufficient for constructing $p$. 
	A t-test is said to be paired if the two data sets consist of pairs of matched units. We utilize pairwise 
	t-tests in the examination of subject completion times.

	\subsection{Pairwise examination}
	\label{subsec:pairs}
	In case of multiple comparisons, researchers can assign an acceptable type I error $\alpha$ 
	(false positive) either to each individual comparison or jointly, across all comparisons. To judge the results along a joint $\alpha$, a significance level 
	$\frac{\alpha}{n}$ can be chosen for each test, with $n$ being the number of pairwise comparisons performed.
	This corresponds to the so-called Bonferroni correction \cite{hochberg1987multiple}.

	\section*{Appendix B: Unattended Experiment Setup} 
	\begin{figure}[h]
	\fbox{\centering
	\includegraphics[height=2.6in,width=1\columnwidth]{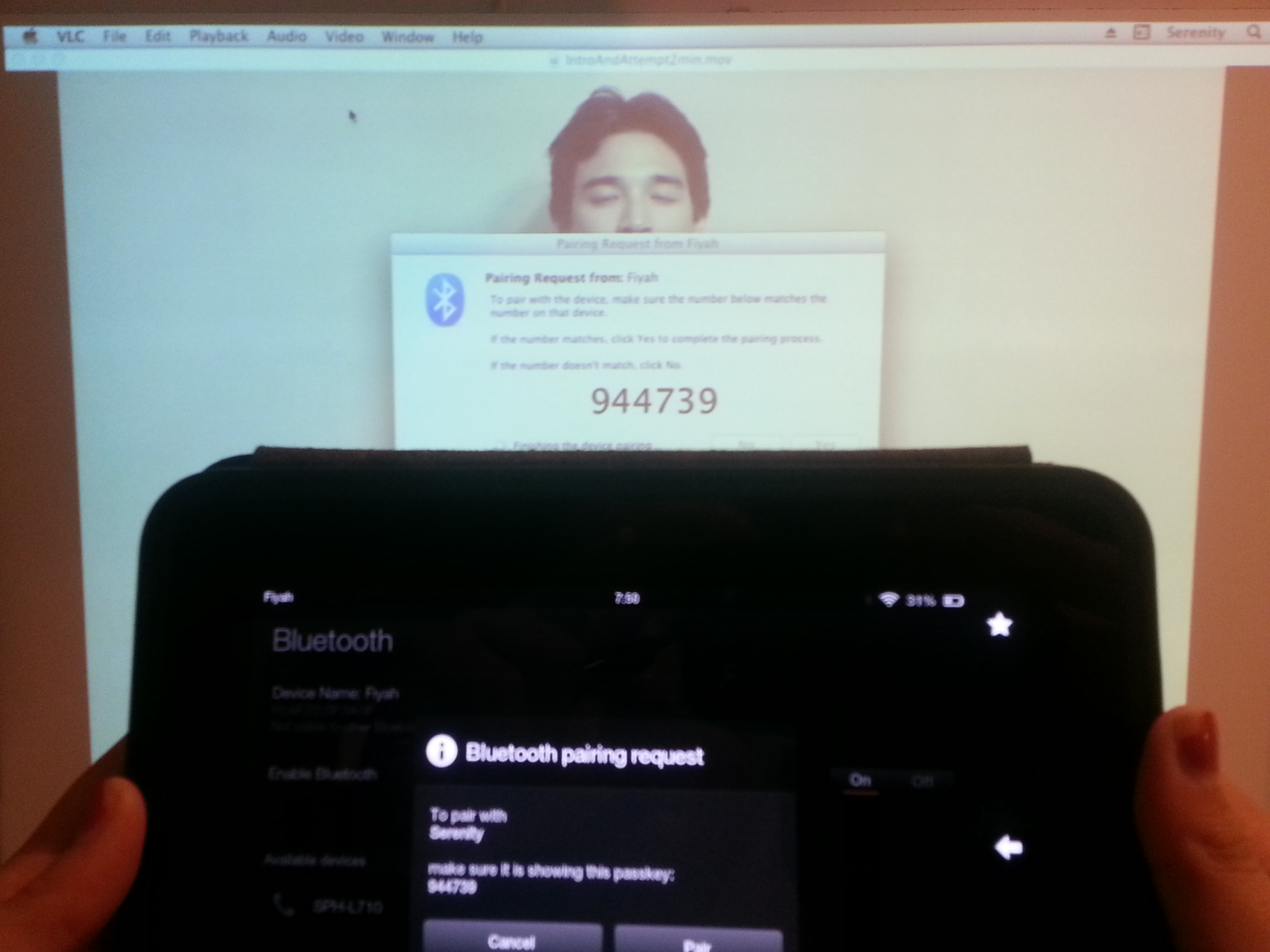}}
	\caption{{\small Bluetooth confirmation screen, from subject's perspective} }
	\label{fig:confirm}
	\end{figure} 

	\begin{figure}[h]
	\fbox{\centering
	\includegraphics[height=2.6in,width=1\columnwidth]{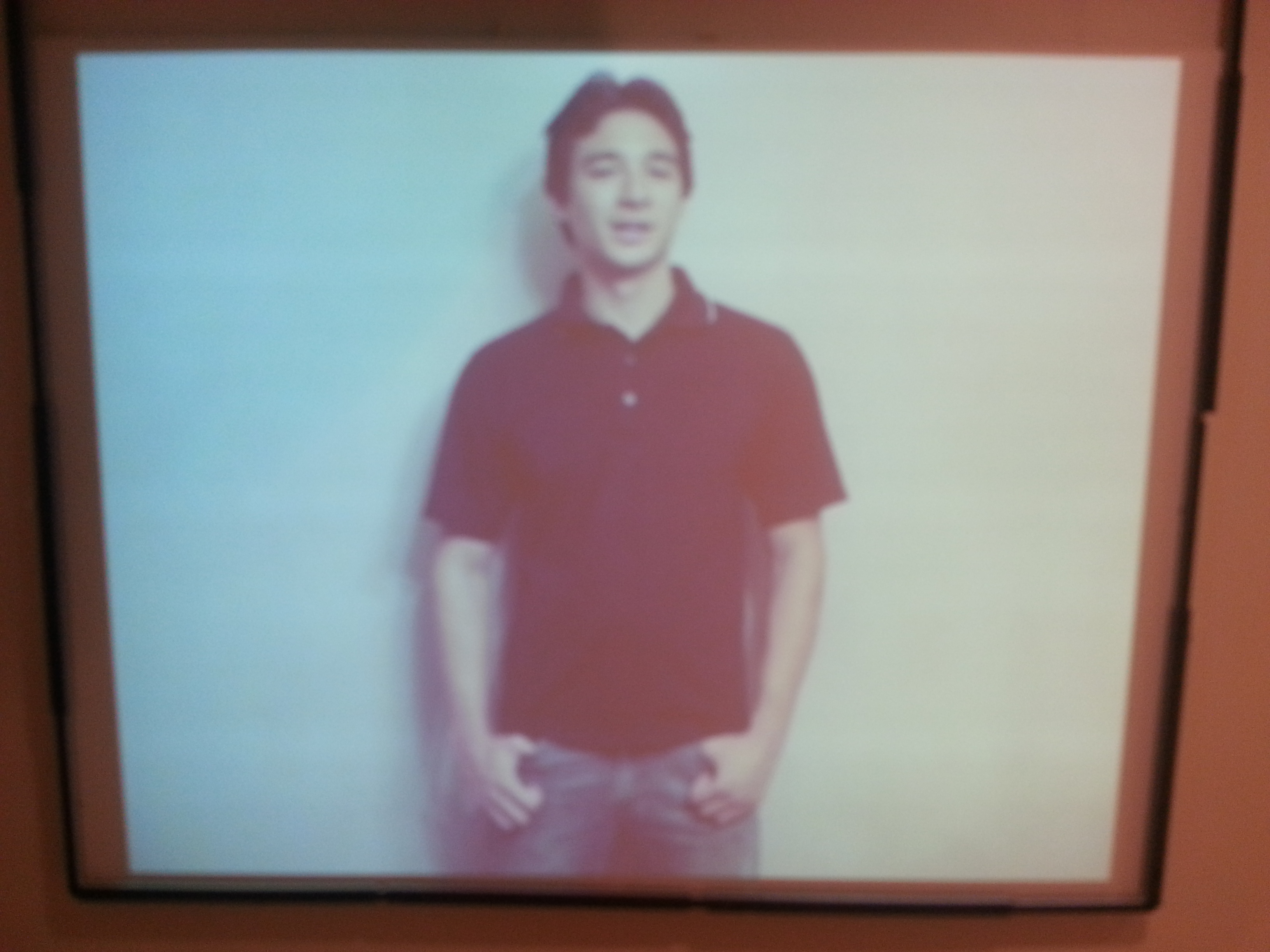}}
	\caption{{\small Experimenter proxy giving video instructions} }
	\label{fig:proxy}
	\end{figure} 

	\begin{figure}[h!]
	\fbox{\centering
	\includegraphics[height=2.6in,width=1\columnwidth]{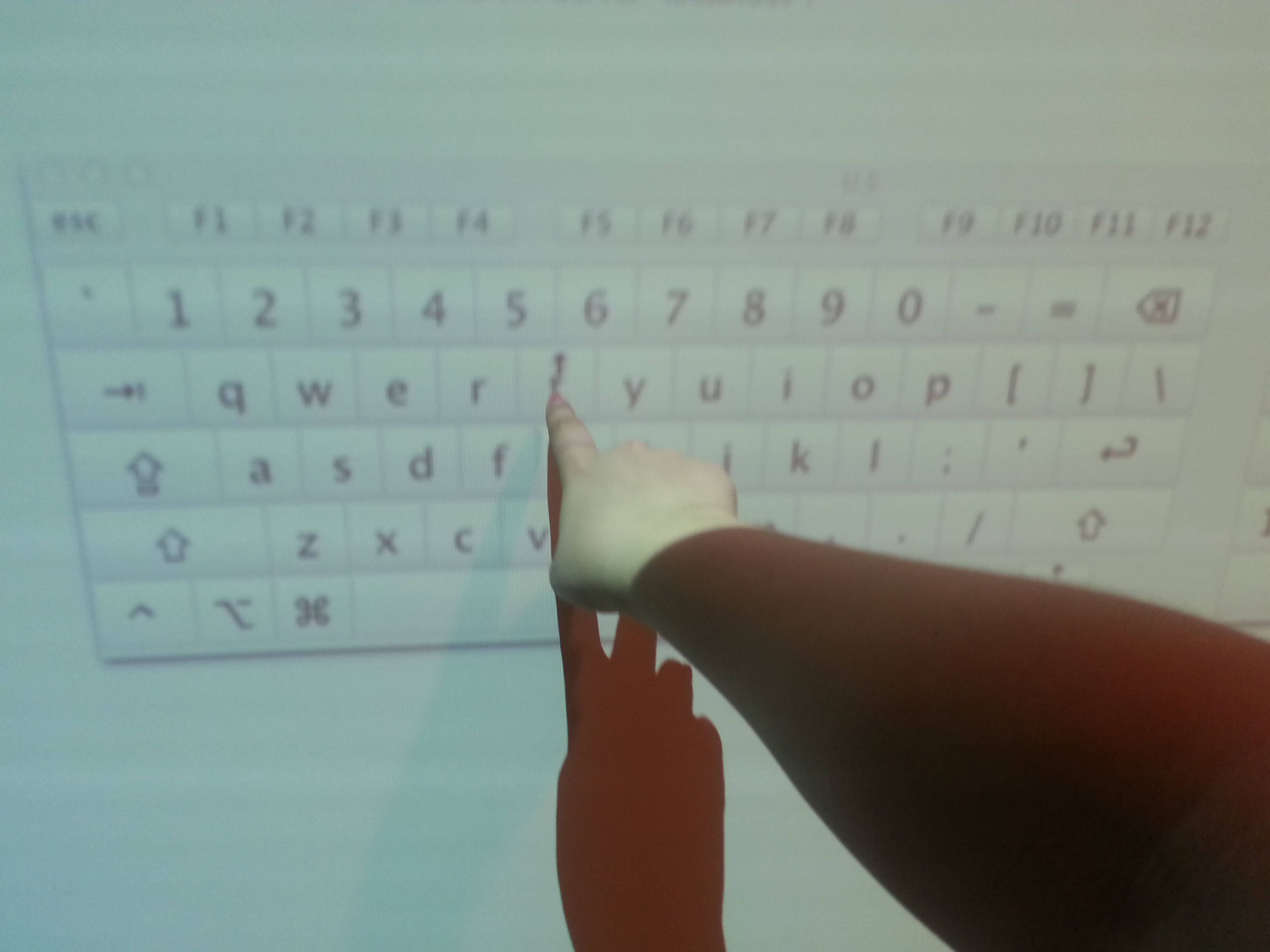}}
	\caption{{\small Subject entering email address on Smartboard} }
	\label{fig:email}
	\end{figure} 

	\begin{figure}[h!]
	\fbox{\centering
	\includegraphics[height=2.6in,width=1\columnwidth]{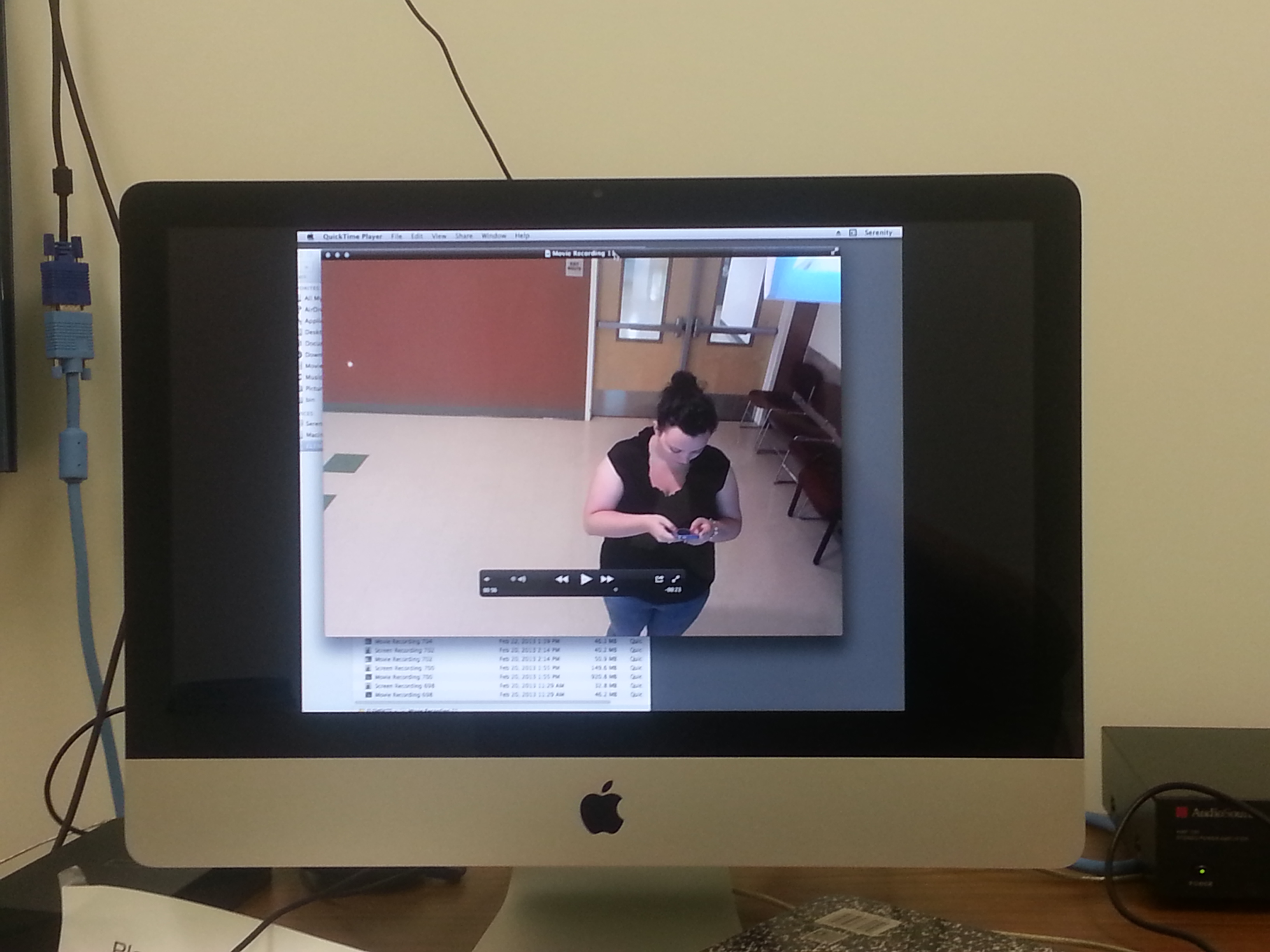}}
	\caption{{\small Post-experimental review of video recordings (separate office)} }
	\label{fig:review}
	\end{figure}


	\end{document}